\documentclass[final,3p,times]{elsarticle}

\usepackage{amssymb}
\usepackage{lipsum}
\usepackage{amsmath,amssymb,bm,mathrsfs}
\usepackage{hyperref}
\usepackage{url}
\usepackage{graphicx}
\usepackage{mathtools}
\usepackage{lipsum}
\usepackage[table]{xcolor}
\usepackage{multirow}
\usepackage{colortbl}
\usepackage{adjustbox}

\begin{document}

\begin{frontmatter}



\title{Non-minimally coupled quintessential inflation}




\author{Seong Chan Park }
\ead{sc.park@yonsei.ac.kr}
\affiliation{organization={Department of Physics, Yonsei University},
           addressline={}, 
            city={Seoul},
            postcode={03722}, 
            state={},
            country={Korea}}
\affiliation{organization={Korea Institute of Advanced Study},
            addressline={}, 
            city={Seoul},
            postcode={02455}, 
            state={},
            country={Korea}}            

\begin{abstract}
We present a unified framework that simultaneously addresses the dynamics of early-time cosmic inflation and late-time cosmic acceleration within the context of a single scalar field non-minimally coupled to gravity. By employing an exponential coupling function and a scalar potential with dual asymptotic plateaus, our model naturally transitions from inflationary dynamics at small field values to a quintessence-like behavior at large field values. We derive the inflationary predictions for the spectral index ($n_s$) and tensor-to-scalar ratio ($r$) in agreement with current observational constraints. For late-time acceleration, the model produces a viable dark energy component with an equation of state $w_\phi$ approaching $-1$ but retaining a measurable deviation that could serve as an observational signature. This work demonstrates that a single theoretical framework can reconcile both early inflation and the late-time accelerated expansion of the Universe.

\end{abstract}




\end{frontmatter}




\section{Introduction}
\label{introduction}

The Standard Big Bang cosmology successfully explains many observed features of the Universe, yet it leaves critical issues unresolved, such as the flatness problem, horizon problem, monopole problem, and the origin of primordial inhomogeneities. Cosmic inflation has emerged as a compelling solution to these challenges~\cite{Starobinsky:1980te, Guth:1980zm, Sato:1981qmu, Linde:1983gd}. The simplest realization of inflation involves a scalar field $\phi$, often referred to as the ``inflaton." Observations from Planck~\cite{Planck:2018jri} and BICEP/Keck experiments~\cite{BICEP:2021xfz} have placed stringent constraints on inflationary parameters, such as the spectral index $n_s$ ($0.958 \leq n_s \leq 0.975$) and the tensor-to-scalar ratio $r \leq 0.036$ (both at $95\%$ confidence level).  

In parallel, late-time observations reveal that the Universe is undergoing accelerated expansion, driven by a dominant dark energy component. This phenomenon has been firmly established through Type Ia supernovae~~\cite{SupernovaSearchTeam:1998fmf,SupernovaCosmologyProject:1998vns}, large-scale structure~~\cite{Daniel:2008et}, and Cosmic Microwave Background (CMB) measurements. While a cosmological constant ($\Lambda$) provides a simple explanation with $w_\Lambda = -1$, its fine-tuning issues motivate alternative models such as dynamical dark energy, particularly quintessence. Recent data from DESI(BAO) indeed suggests an intriguing hint of $w\neq -1$ ~\cite{DESI:2024mwx}:
\begin{align}    \label{eq:DESI}
    w=\begin{cases}
    -0.99^{+0.15}_{-0.13} & \text{DESI BAO}, \\
    -1.122^{+0.062}_{-0.054} & \text{DESI BAO+CMB}, \\
    -0.997 \pm 0.025 & \text{DESI BAO+CMB+PantheonPlus}.
    \end{cases}
\end{align}

Quintessence posits a scalar field with a time-dependent equation of state parameter, distinguishing it from $\Lambda$. See~\cite{Cai:2009zp, Li:2011sd, Bamba:2012cp, Tsujikawa:2013fta} for reviews. A natural question arises: can a single scalar field be responsible for both early-time inflation and late-time dark energy?   Such a unified framework~\cite{Peebles:1998qn, Uzan:1999ch} (also see \cite{Cai:2010zw, Cai:2010kp} ) would not only simplify our theoretical models but also address the fine-tuning challenges inherent in conventional approaches~\cite{deHaro:2021swo}. In this work, we investigate this possibility by constructing a model that employs a single scalar field non-minimally coupled to gravity. Indeed, the non-minimal coupling~\cite{Futamase:1987ua} helps to realize early time inflation~\cite{Park:2008hz, Kallosh:2013hoa, Kallosh:2013yoa} in a wide range of models~\cite{Cheong:2020rao, Jinno:2018jei, Hyun:2022uzc,Hyun:2023bkf,Koh:2023zgn, Jho:2022wxd, Cheong:2023hrj,Cheong:2022ikv, Cheong:2024kum} especially Higgs inflation~\cite{Bezrukov:2007ep}, its critical realization~\cite{Hamada:2014wna,Hamada:2014iga} and $R^2$ generalization ~\cite{Ema:2017rqn, Gorbunov:2018llf, Gundhi:2018wyz, He:2018mgb,Cheong:2019vzl,Cheong:2022gfc} (also see \cite{Cheong:2021vdb} for a review). The model features a carefully designed potential with two asymptotic plateaus, enabling inflation at small field values and late-time cosmic acceleration at large field values. 


This paper is organized as follows: Section 2 introduces the theoretical setup, including the scalar potential and its coupling to gravity. Section 3 discusses the model's inflationary dynamics and its compatibility with observational data. Section 4 examines the late-time acceleration regime, highlighting the observational implications. Finally, Section 5 summarizes the findings and outlines future directions for research.

\section{Model}
In this section, we introduce our model for the early time inflation and the late time acceleration with two asymptotic plateaus. We first introduce the general idea then provide a specific example. 

\subsection{General case}

Our model in Jordan frame is based on the action for a scalar field $\phi$ with a non-minimal coupling with Ricci scalar and a potential
\begin{align}
&S_J =\int d^4 x \sqrt{-g_J} \left(
\frac{M_P^2}{2}\Omega^2(\phi) R_J -\frac{1}{2}(\partial \phi)^2 -V_J(\phi) \right), 
\end{align}
where $\Omega^2(\phi) = 1+K(\phi)$ and $V_J(\phi) = V_0 K(\phi)^2$ to guarantee a plateau when $K(\phi)$ is large.   The Einstein frame is recovered by Weyl transformation : $g_{J,\mu\nu} \to g_{E,\mu\nu}=\Omega^2 g_{J,\mu\nu}$:  
\begin{align}
    S_E=\int d^4x \sqrt{-g_E}\left[\frac{M_P^2}{2}R_E-\frac{1}{2}(\partial h)^2-V_E(h)\right]
\end{align}
where the field $h$ is canonically normalized and $R_J = \Omega^2 (R_E-(3/2)g_J^{\mu\nu}\partial_\mu \log \Omega^2 \partial_\nu \log \Omega^2 +3 \square \log \Omega^2)$. The Einstein-frame potential is given as 
\begin{align}
V_E(h(\phi)) =\frac{V_J(h(\phi))}{\Omega^{4}(h(\phi))}=
\frac{V_0K^2}{(1+K)^2} = V_0 \left(1+K^{-1}\right)^{-2},
\label{eq:potE}
\end{align}
where the Jordan field $\phi(h)$ is understood as an implicit function of the field $h$. The canonical field is explicitly obtained by integrating the relation~\cite{Park:2008hz}
\begin{align}\label{dsdvarphi:eps}
    \frac{dh}{d\phi}
    &=\sqrt{\frac{1+K+\tfrac{3}{2}M_P^2K_{,\phi}^2}{(1+K)^2}}
    =: F(\phi),
\end{align}
where $K_{,\phi}=\partial_\phi K(\phi)$ is understood. 

We focus on a potential that supports both early-time inflation and late-time dark energy, achieved through the presence of two asymptotic plateaus in the small and large field limits. To realize this, we consider $K(\phi)$ as a {\it decreasing} function of the inflaton field, which asymptotically vanishes ($K \to 0$) at large field values. In this framework, early inflation occurs in the small field limit, where the potential is high, while the late-time cosmic acceleration is driven by a small dark energy density in the large field limit, where the potential is low.
\begin{align}
V_E =\begin{cases}
    V_0 (1-2K^{-1})\sim \rho_{\rm inf} &\text{at small field, $K\gg 1$},\\
    V_0 K^2 \sim \rho_{\rm DE}, &\text{at large field, $K\ll 1$}.
\end{cases}
\end{align}
In general, the term $M_P^2 K_{,\phi}^2$ can take on any numerical magnitude. Therefore, a more detailed analysis requires specifying a definite functional form.

\subsection{Exponential non-minimal coupling}

\begin{figure}[t]
    \centering
    \includegraphics[width=0.6\linewidth]{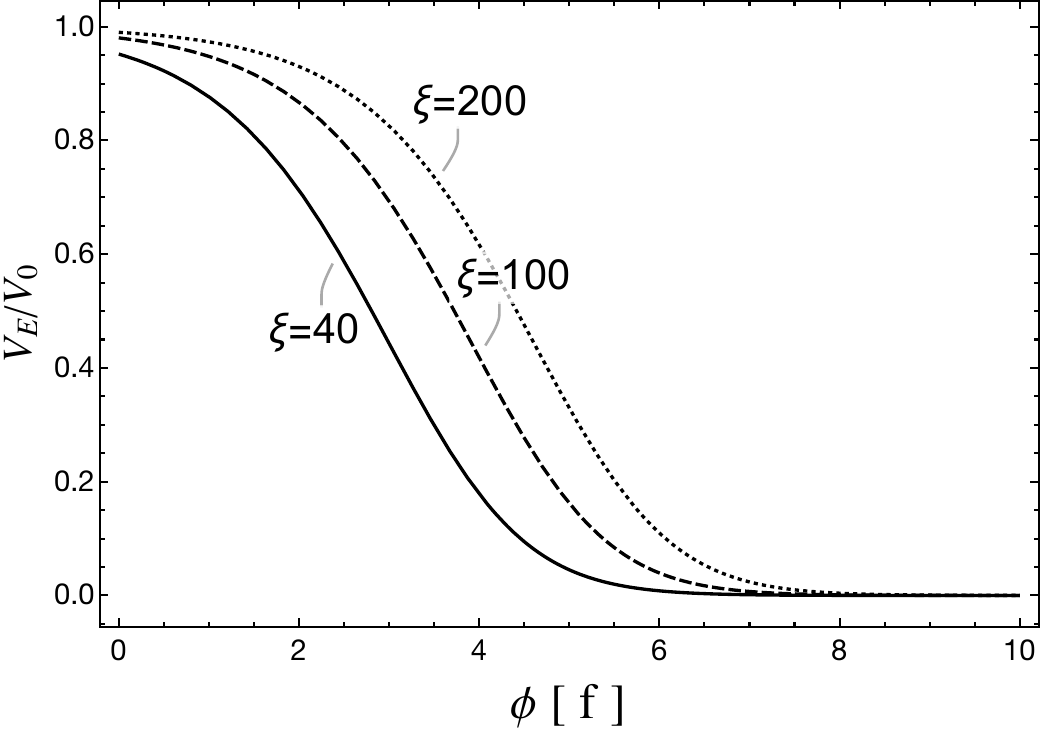}
    \caption{The potential in Einstein frame in terms of $\phi$. We set $\xi=40,100,200$ and $f=1 M_P$ for illustration purpose.}
    \label{fig:potE}
\end{figure}

Partially motivated by a well-studied exponential potential ~\cite{Lucchin:1984yf, Halliwell:1986ja, Burd:1988ss}, we take an exponential type non-minimal coupling that has the desired monotonic decreasing and asymptotically vanishing properties as we described earlier. With this particular $K$, the potential in the Einstein frame is explicitly given as
\begin{align}
    K=\xi e^{-\phi/f}, \quad
    V_E =V_0 \left(1+\frac{e^{\phi/f}}{\xi}\right)^{-2}.
\end{align}
In this work, we will focus on the metric formulation of gravity even though Palatini formulation may have different consequences ~\cite{Jinno:2019und, Cheong:2021kyc, Cheong:2022ikv, Jinno:2018jei}. 
The Einstein frame potential in Eq.~\eqref{eq:potE} has asymptotic behaviors at large field limit ($\phi \ll \xi \ln \xi$) and at small field limit ($\phi \gg \xi \ln \xi$) respectively:
\begin{align}
V_E =\begin{cases}
    V_0 (1-\frac{2}{\xi}e^{\phi/f})& \phi \ll f \ln \xi, \\
    V_0 \xi^2 e^{-2\phi/f} & \phi \gg f\ln \xi,
\end{cases}    
\end{align}
therefore the potential becomes flat at a small field limit but it has a shape of run-away potential at a large field limit as we desired. 
The schematic shape of the potential is shown in the Fig.~\ref{fig:potE}.  Clearly, our model is distinctive from the earlier attempt for scaling solution~\cite{Liddle:1998xm}, and more recent model with a generalized exponential potential~\cite{Geng:2017mic}.

The slow-roll parameters are explicitly given as
\begin{align*}
\epsilon
&=\frac{2M_P^2}{f^2(1+ \xi e^{-\phi/f})+\tfrac{3}{2}\xi^2 M_P^2e^{-2\phi/f}},\\
\eta
&=-\frac{2M_P^2f(1+\xi e^{-\phi/f})}{f^2(1+ \xi e^{-\phi/f})+\tfrac{3}{2}\xi^2 M_P^2e^{-2\phi/f}}. 
\end{align*}

\begin{figure}[t]
    \centering
    \includegraphics[width=0.6\linewidth]{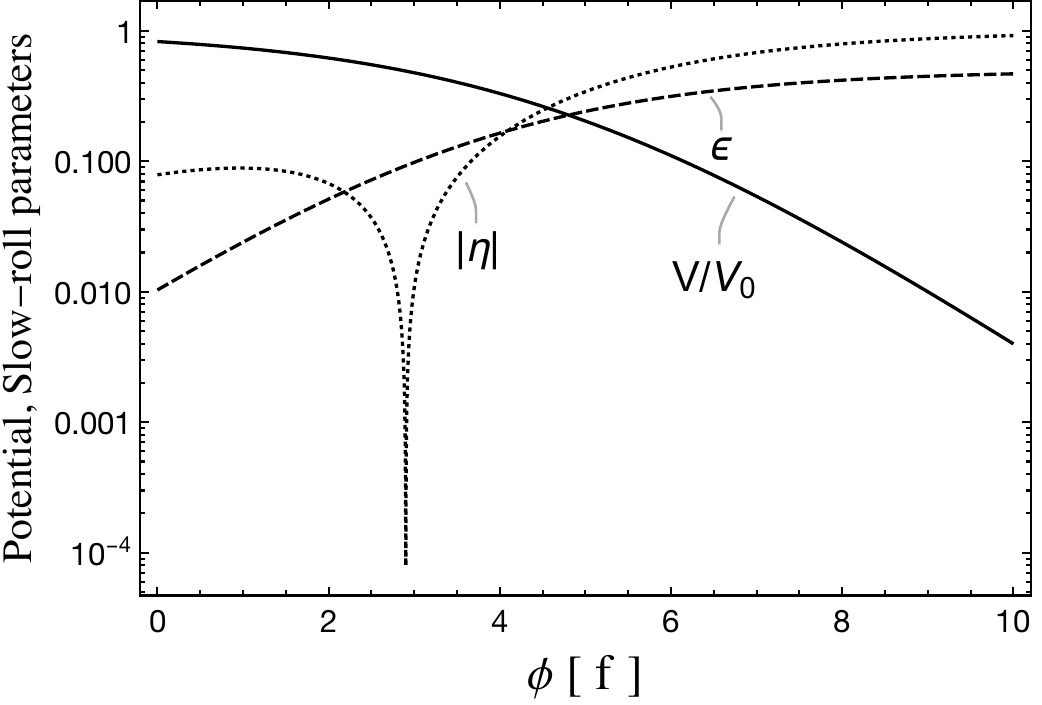}
    \caption{Potential and the slow-roll parameters }
    \label{fig:slow-roll}
\end{figure}

In Fig.~\ref{fig:slow-roll} we depicted the slow-roll parameters $\epsilon$ and $\eta$ in different field values $\phi$ in a small field regime. We set $f=2M_P$ and $\xi=10$ for the illustration purpose.

\section{Early time inflation}

In the vicinity of the origin, $\phi \ll f\ln \xi$ or $K= \xi e^{-\phi/f} \gg 1$,  the potential becomes flat and potentially supports the early time inflation. In this regime (small-field limit), we find an approximate expression:
\begin{align}
    &\Omega^2 \approx K(\phi) \nonumber \\
    &\Rightarrow F^2(\phi) \approx \frac{K+\tfrac{3}{2}M_P^2K_{\phi}^2}{K^2}
=\frac{1}{K} +\frac{3}{2}M_P^2\left((\ln K)_{,\phi}\right)^2. 
\end{align}
 We also simplify $F(\phi)$ as
\begin{align*}
  F(\phi) \approx \sqrt{\frac{1}{\xi e^{-\phi/f}}+\frac{3M_P^2}{2f^2}} \simeq \frac{\sqrt{3/2}M_P}{f}.
\end{align*}

The slow-roll parameters are given in the small field limit $K=\xi e^{-\phi/f} \gg 1$, and $M_P^2 (K')^2 =(M_P/f)^2 K^2 \gg K$:
\begin{align*}
\epsilon
\approx \frac{4}{3K^2}=\frac{4}{3\xi^2 e^{-2\phi/f}},\quad
\eta
\approx -\frac{4}{3K} =-\frac{3}{3\xi e^{-\phi/f}}.
\end{align*}

The number of e-foldings during inflation is calculated in the small field limit:
\begin{align}
N_e \approx \int_{phi_{\text{end}}}^{phi_*} \frac{Fd\varphi}{M_P\sqrt{2\epsilon}} \approx \frac{3\xi}{4}\left(e^{-\phi_*/f}-e^{-\phi_{\rm end}/f}\right) \approx \frac{3}{4}K_*
\end{align}
where $*$ is for the CMB pivot scale where $N_e=(50-60)$. 

\begin{align} \label{ns:eps}
    n_s&=1-6\epsilon_*+2\eta_* \approx 1-\frac{2}{N_e}-\frac{9}{2 N_e^2}, \\
    r&=16\epsilon_{*} \approx \frac{12}{N_e^2},\\
    A_s&=\frac{1}{12\pi^2 M_P^2}\frac{V_E(s_*)^3}{V_{E,s}(s_*)^2}
    \approx \frac{\sqrt{3/2}}{36\pi^2} \left(\frac{M_P}{f}\right)\frac{V_0}{M_P^4} N_e^2.
\end{align}

Requesting $60$ e-foldings,  we determine the inflationary predictions for the spectral index and the tensor-to-scalar ratio
\begin{align}
    n_s =0.965, r=0.003.
\end{align} 
It is noticed that these results exactly coincide with the results of the non-minimally coupled model of inflation with a general power law (See. Ref. \cite{Park:2008hz} in detail.) Finally, we determine the parameter $V_0$ by requesting the CMB normalization condition 
 $\ln{\left(10^{10}A_s\right)}=3.044\pm 0.014$ (TT,TE,EE+lowE+lensing)~\cite{Planck:2018jri}, leads:
\begin{align} 
    \frac{V_0}{f M_P^3} \approx 
    (1.67-1.71) \times 10^{-10} \times \left(\frac{N_e}{60}\right)^{-2}.
    \label{eq:CMB}
\end{align}

\section{Late time Dark energy}

In the large field limit, $K\ll 1$ or $\phi \gg f\ln \xi$, the Einstein frame becomes indistinguishable from the Jordan frame as $\Omega^2 \approx 1$ and $F\approx 1$ or $h\approx \phi$. The potential is a run-away type so that the energy density will go down as time goes on:
\begin{align}
V_E =V_0 \xi^2 e^{-2\phi/f}.
\end{align}
Note that this is nothing but the quintessence potential considered in ~\cite{Barreiro:1999zs}, which has been extensively studied.  

The slow-roll parameters are approximately constant in this limit:
\begin{align}
    \epsilon \simeq \frac{2M_P^2}{f^2},
    \eta \simeq \frac{4M_P^2}{f^2}.
\end{align}
The slow-roll approximation is valid when $f\gg M_P$. We estimate the size of the kinetic term using the slow-roll approximation
\begin{align}
    \dot{\phi}^2 \approx \left(-\frac{V'_E(\phi)}{3 H}\right)^2 
    \approx \frac{\Omega_{\phi}M_P^2V_E'^2}{3 V_E} = \frac{2}{3}V_E \Omega_{\phi} \epsilon,
\end{align}
which leads to the equation of state as 
\begin{align}
    w_\phi = \frac{\tfrac{1}{2}\dot{\phi}^2-V_E(\phi)}{\tfrac{1}{2}\dot{\phi}^2 + V_E(\phi)} \approx  -1+\frac{2 \Omega_{\phi}}{3}\epsilon \approx -1 +\frac{4 \Omega_\phi}{3}\left(\frac{M_P}{f}\right)^2.
\end{align}

\begin{figure}[t]
    \centering
    \includegraphics[width=0.6\linewidth]{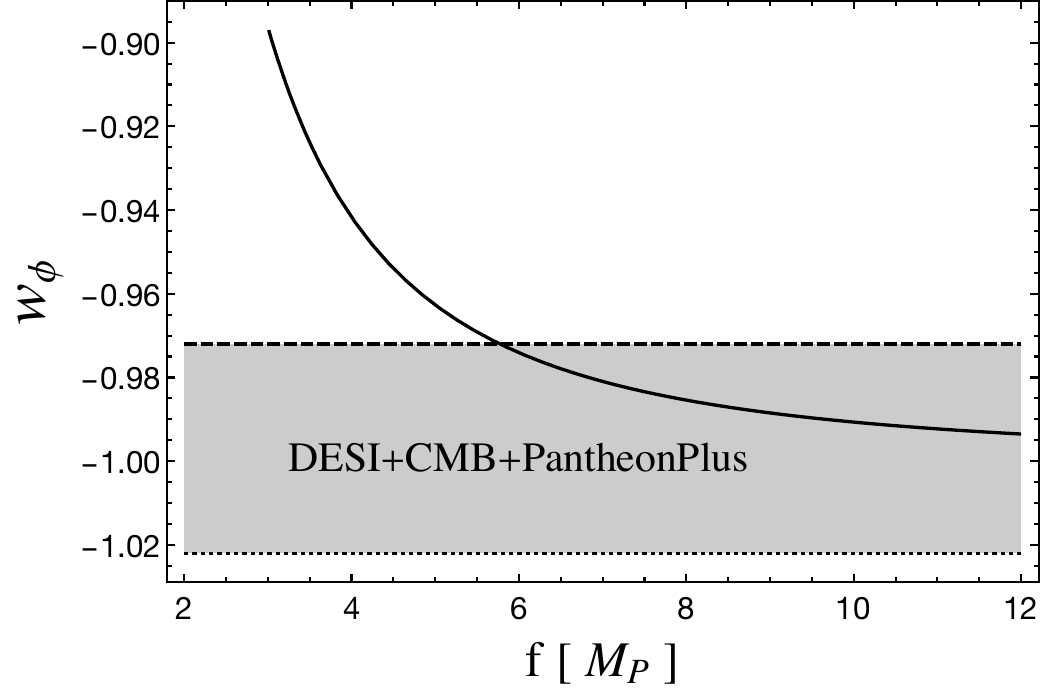}
    \caption{The equation of state \( w_\phi \) is analyzed for $f \in (1, 10)M_P$. To fully account for the dark energy component in the current Universe, we set $ \Omega_\phi = 0.7$. The DESI BAO+CMB+PantheonPlus region~\cite{DESI:2024mwx} is shaded. 
}
    \label{fig:wphi}
\end{figure}

Figure~\ref{fig:wphi} shows the equation of state for our model under the slow-roll approximation. For comparison, we also include recent results from DESI BAO+CMB+PantheonPlus~\cite{DESI:2024mwx}. In the large $f$ limit, the predicted value approaches $-1$, consistent with the cosmological constant; however, deviations from $-1$ are possible. Such deviations would serve as a distinctive signal of the quintessence field $\phi$. Notably, the model remains fully compatible with recent data for $f\gtrsim 7$.

We now discuss the trajectory of quintessence field $\phi$ in a large-field limit with respect to the number of e-foldings $N$. We set the current universe by $N=0$ so that $N<0$ in an earlier time.
\begin{align} \label{Evolutions:eps}
    &\frac{d\phi}{dN}  =\frac{d\phi}{dt}\frac{dt}{dN}\simeq-\frac{V_{E,\phi}}{3H^2}\simeq -\frac{\Omega_{\phi}M_P^2V_{E,\phi}}{V_E}=\frac{2M_P^2\Omega_\phi}{f}  \nonumber \\ 
    &\Longrightarrow \phi(N)\simeq 
\phi_i+\frac{2M_P^2\Omega_\phi}{f}(N-N_i) .
\end{align}
where $\phi_i \equiv \phi(N=N_i)$ set at an earlier moment with $N_i<0$. The energy density of quintessence $\rho_\phi$ with respect to $N$ is given by
\begin{align} \label{analyticrhovarphi:eps}
    \rho_{\phi} (N) \simeq V_E(\phi(N))\simeq \frac{V_0 \xi^2}{e^{2\phi(N)/f}}.
\end{align}
Finally, matching the Hubble parameter at the current Universe  $N=0$ tells us
\begin{align}
V_E \approx 3M_P^2 H^2\Omega_\phi 
\Rightarrow \phi_0=\phi(N=0) \approx \frac{f}{2} \ln \left[\frac{V_0\xi^2}{3M_P^2 H^2\Omega_\phi}\right].
\end{align}
Taking $\rho_{\rm crit} =3M_P^2 H^2 \sim m_\nu^4$ with the neutrino mass scale $m_\nu \sim 10^{-3}$ eV and the CMB normalization condition $V_0 \sim 10^{-10}fM_P^3$ given in Eq.~\eqref{eq:CMB},   we read  \begin{align}
    \phi_0 \simeq \frac{f}{2}\ln \left[\frac{10^{-10}fM_P^3\xi^2}{\Omega_\phi m_\nu^4}\right].
\end{align}

\section{Summary and conclusions}

In this work, we have proposed a unified theoretical framework that effectively addresses both early-time inflation and late-time dark energy using a single scalar field non-minimally coupled to gravity. By employing a carefully constructed potential with dual asymptotic plateaus, we demonstrated how the model transitions seamlessly from inflationary dynamics at small field values to a quintessence-like regime at large field values. 

The model achieves compatibility with current observational constraints on inflation, producing predictions for the spectral index ($n_s \approx 0.965$) and tensor-to-scalar ratio ($r \approx 0.003$) that align with Planck and BICEP/Keck data. Furthermore, the late-time dynamics naturally yield a dark energy equation of state approaching $w_\phi \to -1$, with measurable deviations providing a potential observational signature of the underlying quintessence mechanism. 

Our results highlight the potential for a single scalar field to unify the seemingly disparate phenomena of early and late cosmic acceleration within a consistent theoretical framework. This approach not only addresses the fine-tuning challenges inherent in conventional models but also opens avenues for new observational tests of scalar-field dynamics across cosmic epochs. Future work will focus on exploring the implications of this model in varying cosmological scenarios and its sensitivity to alternative formulations of gravity.

Finally, we comment on the cosmological constant problem. Within our framework, the problem can be reformulated as $\rho_{\varphi,0} = V_0 \xi^2 / e^{2\phi_0 / f} \ll M_P^4$. This condition can be satisfied by assuming a large field value for $\phi_0$, specifically $\phi_0 \sim \mathcal{O}(100)f$.

\section*{Acknowledgements}
I am deeply grateful to Yi-fu Cai and Masahide Yamaguchi for their inspiring discussions. This work was supported by the National Research Foundation of Korea (NRF) grants funded by the Korean government (MSIT) under project numbers RS-2023-00283129 and RS-2024-00340153.

\appendix




\bibliographystyle{unsrt}
\bibliography{refs}

\end{document}